\documentclass[prb,preprint,showpacs,preprintnumbers,amsmath,amssymb]{revtex4}
\usepackage{graphics}
\begin{document}

\title{Self-consistent equilibrium of a two-dimensional electron system with a reservoir
in a quantizing magnetic field: Analytical approach}

\author{V. G. Popov}
\affiliation{Institute of Microelectonics Technology RAS,
Chernogolovka 142432, Russia.}

\date{\today}

\begin{abstract}
An analytical approach has been developed to describe grand
canonical equilibrium between a three dimensional (3D) electron
system and a two dimensional (2D) one, an energy of which is
determined self-consistently with an electron concentration. Main
attention is paid to a Landau level (LL) pinning effect. Pinning
means a fixation of the LL on a common Fermi level of the 2D and
the 3D systems in a finite range of the magnetic field due to an
electron transfer from the 2D to the 3D system. A condition and a
start of LL pinning has been found for homogeneously broadened
LLs. The electronic transfer from the 3D to the 2D system controls
an extremely sharp magnetic dependency of an energy of the upper
filled LL at integer filling of the LLs. This can cause a
significant increase of inhomogeneous broadening of the upper LL
that was observed in recent local probe experiments.
\end{abstract}

\maketitle

\section{\label{sec:int}INTRODUCTION}
A model of grand canonical equilibrium of a 2D system with a
reservoir was proposed to explain the integer quantum Hall effect
(IQHE) more than twenty years ago.\cite{Baraff} This so-called
reservoir hypothesis can describe main features accompanied to the
IQHE such as plateaus in a Hall resistance,\cite{Baraff} the
Shubnikov-de Hass effect\cite{Raymond} and oscillations of a
magnetization and a thermoelectric power.\cite{Zawadskii} Key
points of the reservoir model are following: a 2D chemical
potential is fixed by the reservoir and it doesn't depend on the
magnetic field, a carriers concentration and 2D subband energies
oscillate in the magnetic field. There was a wide discussion about
a relevance of the hypothesis to conventional samples where the
IQHE observed and it is still not clear. But there are structures
where a reservoir is created nearby to a 2D system and one should
take in to account the carriers exchange. These are tunnel and
resonant tunnelling diodes.\cite{Mizuta} For these devices it was
shown that a carriers transfer between 2D and 3D systems takes
place and a 2D subband energy oscillates while a magnetic field
perpendicular to the 2D plane sweeps.\cite{Eaves} Moreover the 2D
subband energy oscillations can be so strong that a pinning effect
of a partial filled Landau level (PFLL) takes
place.\cite{Skolnick,Takagaki} Usually authors employ
self-consistent calculations to get magnetic dependencies of a
carriers concentration and a 2D subband energy and to my knowledge
there is no any thorough analytical investigation of the LL
pinning effect. In this paper I would like to consider thoroughly
pinning of a homogeneously broadened PFLL with aims to draw
analytical expressions, to find a PFLL pinning condition and to
make other estimations.

Recently a new local probe technique so-called a subsurface charge
accumulation (SCA) imaging has been developed by S. H. Tessmer, R.
C. Ashoori et al\cite{Tessmer} to investigate a local
compressibility of a 2D electron system (2DES). Extraordinarily
sensitive to a magnetic field features have been revealed in the
images at the magnetic fields close to the integer quantum Hall
state of the 2DES. The SCA is measured as a charge on a tip
induced by an applied ac voltage between the metal tip and a
contact to the 2DES. A frequency of the signal is chosen quite low
to provide the charge accumulation without delay. In this case the
SCA is proportional to a capacitance between the tip and the 2DES
that is determined by the local compressibility of the 2DES. In
other words all applied ac voltage drops between the tip and the
2DES, namely, Fermi levels of the 2DES and the contact are equal.
This means equilibrium between the 2DES and the contact. Hence at
the SCA image analysis one should takes in to account effects of
this equilibrium.

An analytic approach of cause requires some simplifications but
the all made assumptions are appeared to be relevant to real
tunnel structures with 2DESs. So I start in Sec.~\ref{sec:mod}
with a $\delta$-profile of a quantum well that allows me to avoid
many subbands in a 2DES. As usual an electron reservoir is
considerably away from the 2DES that permits to neglect a
variation of a width of a localized wave function and to use only
the Poisson equation in Sec.~\ref{sec:pin} instead of
self-consistent calculations of the Poisson and the Shr\"{o}dinger
equations. The next assumption is in homogeneous broadening of the
LLs. In this case a shape of the LL can be described analytically
for a large LL number.\cite{Ando} Relevance of this approach is
also discussed in Sec.~\ref{sec:pin}. In Sec.~\ref{sec:res} the
subband energy oscillations in the magnetic field have been
considered when PFLL pinning takes place and a starting value of
the magnetic field has been determined for PFLL pinning. In
Sec.~\ref{sec:concl} I conclude this paper.

\section{\label{sec:mod}MODEL OF STRUCTURE}
\begin{figure}
\includegraphics{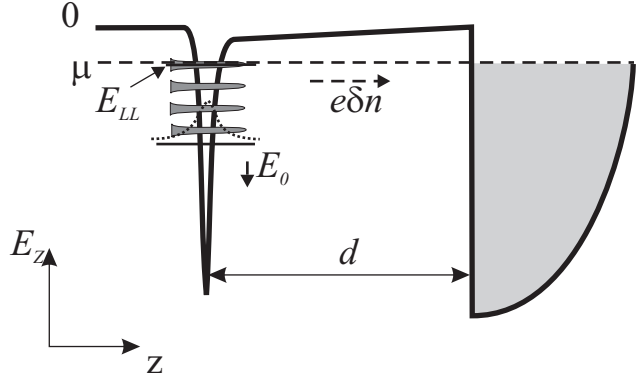} \caption{\label{fig:model} Potential
profile of the model structure with the Landau staircase and the
2D subband level $E_0$ in the quantizing magnetic field. Dashed
arrows show the charge transfer and the 2D subband shift at the
magnetic field increase while the PFLL is pined at the common
Fermi level.}
\end{figure}

Let us consider a quantum well (QW) separated from a metal surface
with a spacer barrier of thickness $d$ (see Fig.~1). To simplify
calculations a QW potential profile is supposed to be of the
following form:
\begin{equation}\label{delta}
    U(z)=-\alpha\,\delta(z).
\end{equation}
It is well known\cite{quant} that such QW has only one localized
state at the energy $E_0=-m\alpha^2/2\hbar^2$ that is also valid
in this situation because I suppose
\begin{equation}\label{width}
    d\gg\hbar^2/m\alpha.
\end{equation}
Here $2\hbar^2/m\alpha$ is an effective width of the localized
wave function; $m$ is an effective mass of the electron and
$\hbar$ is the Planck's constant. It should be mentioned that a
depletion effect inherent to a metal-semiconductor junction is out
of the consideration and the potential profile of the structure at
a magnetic field is shown in Figure~\ref{fig:model}. As the next
step I consider a dirty metal so that a magnetic field applied
alone z axis doesn't influence on the metal density of states
(DOS) and creates the Landau levels (LLs) in to the 2DES. This
approximation is quite usual for semiconductor devices especially
if one considers metal contacts. A temperature is supposed to be
zero.

\section{\label{sec:pin}PINNING OF LANDAU LEVEL}
A magnetic field applied alone the z axis quantizes the 2D states
in to the Landau levels (LLs). It is well known that a Fermi
energy of a 2DES oscillates in a quantizing magnetic field for a
constant electron concentration of a 2DES. In my case the 2DES
Fermi level is fixed by the metal thus a constant concentration
$n$ is not relevant. Carried out in Ref.~\onlinecite{Eaves}
thorough calculations show that $E_0$ and $n$ are oscillating
functions of the magnetic field. My task here is to demonstrate
that there is a situation when the subband energy $E_0$ changes in
the magnetic field in such manner that the PFLL is almost fixed or
pined on the common Fermi level of the junction. Let me start from
a definition of a 2D density of states (DOS) $G(\varepsilon, B)$:
\begin{equation}\label{con}
    n=\int_{0}^{\mu - E_0}\,G(\varepsilon,B)d\varepsilon,
\end{equation}
where $\mu$ is the common Fermi energy. Applying a variation
$\delta B$ of the magnetic field one should expect the following
variation of the electron concentration:
\begin{equation}\label{dn}
    \delta n=-G(\mu - E_0,B)\,\delta E+\int_0^{\mu-E_0}\frac{\partial G(\varepsilon,B)}{\partial
    B}d\varepsilon\,\delta B.
\end{equation}
Now I would like to substitute on the place of the $\delta B$ a
variation of the cyclotron energy $\delta\varepsilon_c = \hbar e
\delta B/m$ and consider a derivative $\partial G(\varepsilon,
B)/\partial\varepsilon_c$. To determine it's value let me consider
the DOS in detail. This value is found as a sum of spectral
functions of the broadened LLs:
\begin{equation}\label{dos}
    G(\varepsilon,B)=\sum_{i=0}^\infty\beta\varepsilon_c\,\varphi_i(\varepsilon,B).
\end{equation}
Here $\beta\varepsilon_c$ is a degeneracy of the LL per unit area
of the 2DES ($\beta = m/\pi\hbar^2$ is the DOS at zero magnetic
field). As for the spectral function it has the following
dependence on $\varepsilon_c$:
\begin{equation}\label{shape}
\varphi_i (\varepsilon , B)=
\frac{1}{\Gamma}\,\psi\!\!\left[\left(\frac{\varepsilon-(i+1/2)\varepsilon_c}{\Gamma}\right)^2\,\right].
\end{equation}
$\Gamma$ is a width of the LL, $i$ is a LL number, the function
$\psi$ describes a shape of the LL. Here I neglect a LL spin
splitting and an electron-electron interactions inside the 2DES.

A shape of the LLs was a subject of an intensive study during the
last three decades.\cite{Ando,Raikh,Benedict} The complexity of
the problem arises from the fact that in the absence of the
disorder the energy spectrum is discrete. As a result, the
self-energy of an electron appears to be real in any finite order
of the perturbation theory. Therefore, obtaining of a finite width
of the LL requires summation of the entire diagram expansion. It
was demonstrated\cite{Ando,Raikh} that such a summation is
possible when the number of the LL is large. The simplifications,
arising in this limit, are different in the case of a short-range
and a smooth disorder. In the former case the correlation radius
$R_c$ of the disorder less than the magnetic length $\ell$ and
only a subsequence of diagrams without self-intersections
contributes to the self-energy, or, in other words, the
self-consistent Born approximation\cite{Ando} becomes
asymptotically exact. The shape of the LL in this case is close to
semielliptical. For a smooth disorder, with $\ell \ll R_c \ll R_L
\approx i^{1/2}\ell$ ($R_L$ is the Larmour radius) the LL shape
had been found Gaussian. In this case the random potential can be
renormalized to effective one with $R_c \geq R_L$ and hence it
becomes a long-range potential.\cite{Raikh} In the case of the
long-range disorder $R_c \gg R_L$ the semiclassical approach is
valid when the disorder simply modulates the subband energy $E_0$
and LL staircase follows this modulation. This means an
inhomogeneous broadening of the LL and it will be considered
somewhere else. Thus only the short-range disorder leads to the
homogeneous LL broadening.

As for the model under consideration the correlation radius $R_c$
could not be longer than the distance between the 2DES and the
metal $d$ because of the metal screening\footnote{Strictly
speaking this estimation is valid in the case of an electrostatic
disorder formed by electric field modulations. Besides the
electrostatic there is also a structural disorder originated from
QW width or layer thickness modulations and QW depth or layer
composition those. In this case $R_c$ is determined by a
correlation radius of these modulations $R_m$ that depends on
technological parameters. In this case to provide homogeneous LL
broadening one needs satisfy $\ell \gg R_m$.}. Hence the
short-range disorder approximation is adequate when $\ell > d$. In
this case the LL spectral function can be written as
\begin{equation}\label{spectral}
    \varphi_i(\varepsilon,B)=\frac{2}{\pi\Gamma}\left[1-\left(\frac{\varepsilon-(i+1/2)\varepsilon_c}
    {\Gamma}\right)^2\right]^{1/2},
\end{equation}
where $\Gamma= \gamma\varepsilon_c^{1/2}$, $\gamma =
(2\hbar/\pi\tau)^{1/2}$ and $\tau$ is a relaxation time at $B =
0$.\cite{Ando}

Other mechanisms of LL broadening such as phonon-electron,
electron-electron interactions are negligible because the
temperature is zero. LL broadening caused by electron tunnelling
between the 2DES and the metal is possible but also negligible due
to a very low tunnel transparency in accordance with
Eq.~(\ref{width}).

Thus having the LL shape determined, one can make the next step.
From Eqs.~(\ref{dn}) and (\ref{dos}) one can get:
\begin{equation}\label{dn1}
    \delta n=-G_F\,\delta E_0 +
    \int^{\mu-E_0}_0\sum^\infty_{i=0}\left(\beta\,\varphi_i(\varepsilon,\varepsilon_c)+
    \beta\varepsilon_c\frac{\partial\varphi_i(\varepsilon,\varepsilon_c)}{\partial\varepsilon_c}\right)
    d\varepsilon\,\delta\varepsilon_c.
\end{equation}
Here I have $G_F = G(\mu - E_0, B)$. In accordance with equations
(\ref{con}) and (\ref{dos}) one can simplify the equation in to
the following:
\begin{equation}\label{dn2}
    \delta n = -G_F\,\delta E_0 + \frac{n}{\varepsilon_c}\,\delta\varepsilon_c +
    \int^{\mu-E_0}_0\sum^\infty_{i=0}\beta\varepsilon_c
    \frac{\partial\varphi_i(\varepsilon,\varepsilon_c)}{\partial\varepsilon_c}\,d\varepsilon\,\delta\varepsilon_c.
\end{equation}
According to Eq.~(\ref{spectral}) the derivative in the last term
of this equation can be expressed as:
\begin{equation}\label{der}
    \frac{\partial\varphi_i(\varepsilon,\varepsilon_c)}{\partial\varepsilon_c}=-\frac{\varphi_i(\varepsilon,\varepsilon_c)}
    {2\varepsilon_c}-\left(\frac{i+1/2}{2}+\frac{\varepsilon}{2\varepsilon_c}\right)
    \frac{\partial\varphi_i(\varepsilon,\varepsilon_c)}{\partial\varepsilon},
\end{equation}
or
\begin{equation}\label{der1}
    \frac{\partial\varphi_i(\varepsilon,\varepsilon_c)}{\partial\varepsilon_c}=-\left(\frac{i+1/2}{2}\right)
    \frac{\partial\varphi_i(\varepsilon,\varepsilon_c)}{\partial\varepsilon}-\frac{1}{2\varepsilon_c}
    \frac{\partial[\varepsilon\,\varphi_i(\varepsilon,\varepsilon_c)]}{\partial\varepsilon}.
\end{equation}
Now one can easy perform integration in the last term of
Eq.~(\ref{dn2}) and get the following:
\begin{equation}\label{dn3}
    \delta n = -G_F\,\delta
    E_0+\frac{n}{\varepsilon_c}\,\delta\varepsilon_c-\sum^\infty_{i=0}\beta\varepsilon_c\left(\frac{\mu-E_0}{2\varepsilon_c}
    +\frac{i+1/2}{2}\right)\,\varphi_i(\mu-E_0,\varepsilon_c)\,\delta\varepsilon_c,
\end{equation}
where I have already used that $\Gamma \ll \varepsilon_c$ and set
$\varphi_i(0, \varepsilon_c) = 0$. Following this approximation
further one can get from Eq.~(\ref{dos}) that $G_F =
\beta\varepsilon_c\,\varphi_N(\mu - E_0, \varepsilon_c)$ ($N$ is a
number of the PFLL). In this case the Eq.~(\ref{dn3}) is modified
to the following:
\begin{equation}\label{dn4}
    \delta n =-G_F\,\delta
    E_0+\frac{n}{\varepsilon_c}\,\delta\varepsilon_c-\left(\frac{\mu-E_0}{2\varepsilon_c}+\frac{N+1/2}{2}\right)
    G_F\,\delta\varepsilon_c.
\end{equation}

The next step is to find a relation between $\delta n$ and $\delta
E_0$. The rigorous way is to solve self-consistently the Poisson
and Schr\"{o}dinger equations but I have inequality~(\ref{width})
that allows me to neglect the variations of the wave function and
QW widths in compare with that of the QW depth. The depth
variation can be easy evaluated from the variation of the electric
field between the metal and the 2DES. In this case the subband
level $E_0$ follows the QW depth. Hence one can get the following:
\begin{equation}\label{dE0}
    \delta E_0=e^2\,\delta n/C_0.
\end{equation}
$C_0$ is a geometric capacity per unit area, i.e., $C_0 =
\kappa/d$, where $\kappa$ is a permittivity of the structure
material\footnote{It is important to note that such simple
expression of $C_0$ is a following of the homogeneous model. In
the case of inhomogeneous LL broadening one should take into
account an in-plane charge exchange that leads to a coordinate
dependency of $C_0$ and in particularly it's dependency upon the
local value of $G_F$. The Eq. (\ref{dE0}) therewith is valid
locally.}. Combining Eqs.~(\ref{dn4}) and (\ref{dE0}) one can
determine the derivative $dE_0/d\varepsilon_c$ as:
\begin{equation}\label{derE0}
    \frac{dE_0}{d\varepsilon_c}=\frac{2e^2n-[\mu-E_0+(N+1/2)\varepsilon_c]G_Fe^2}{2e^2\varepsilon_cG_F+2\varepsilon_cC_0}.
\end{equation}
According to this equation the derivative of a central energy of
the PFLL $dE_{\rm{LL}}/d\varepsilon_c$ can be calculated as
\begin{equation}\label{derELL}
    \frac{dE_{\rm{LL}}}{d\varepsilon_c}=N+\frac{1}{2}+\frac{dE_0}{d\varepsilon_c}=\frac{2e^2n+2\varepsilon_cC_0(N+1/2)+
    [(N+1/2)\varepsilon_c-\mu+E_0]e^2G_F}{2\varepsilon_cG_F+2\varepsilon_cC_0}.
\end{equation}
PFLL pinning takes place when the derivative is zero or negligibly
small, i.e., $dE_{\rm{LL}}/d\varepsilon_c \ll 1$. Hence the
pinning condition is the following:
\begin{equation}\label{pincon}
    G_F \gg n/\varepsilon_c+NC_0/e^2,
\end{equation}
here I have used that in the case of the large $G_F$ the Fermi
level is very close to the LL center and the following inequation
is justified
$\mu-E_0-(N+1/2)\varepsilon_c<\Gamma\ll\varepsilon_c$. Also the $N
\gg 1$ is supposed.

\section{\label{sec:res}RESULTS AND DISCUSSION}
\begin{figure}
\includegraphics{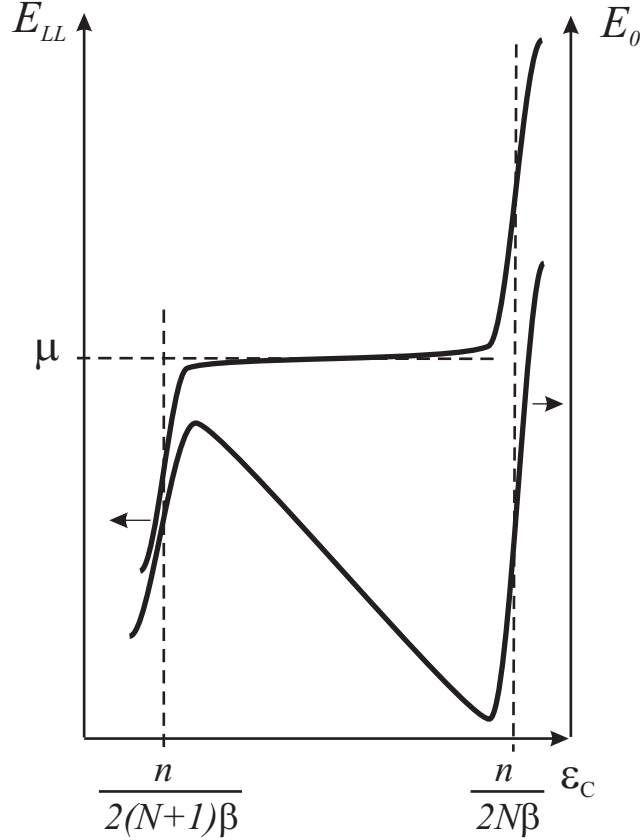}
\caption{\label{fig:oscil} Magnetic dependencies of the subband
energy $E_0$ and the central energy $E_{\rm{LL}}$ of the $N$-th
PFLL.}
\end{figure}

Since the PFLL is pined on the Fermi level the subband energy
$E_0$ should have a magnetic dependence. From Eqs. (\ref{derE0}),
(\ref{pincon}) one can get a quite expectable result for this
dependence:
\begin{equation}\label{pinE0}
    dE_0/d\varepsilon_c=-(N+1/2).
\end{equation}
This means that under the pining condition a basement of the LL
staircase $E_0$ shifts down on the energy scale and this shift
compensates the PFLL energy increase due to the increase of the LL
staircase step with the magnetic field (see Fig.~1). For integer
filling of the LLs $G_F$ is zero and from Eqs.~(\ref{derE0}) and
(\ref{derELL}) one can get:
$dE_0/d\varepsilon_c=e^2n/(\varepsilon_cC_0)$ and
$dE_{\rm{LL}}/d\varepsilon_c=N+1/2+e^2n/(\varepsilon_cC_0)$. In
the case of integer filling factor the electron concentration can
be expressed as $n = N\beta\varepsilon_c$ then one can get
\begin{equation}\label{E0jump}
    dE_0/d\varepsilon_c=e^2 \beta N/C_0,
\end{equation}
and
\begin{equation}\label{ELLjump}
    dE_{\rm{LL}}/d\varepsilon_c=(1+e^2\beta /C_0)N-1/2.
\end{equation}
Here I suppose the $N$-th LL as just emptied. Let me make
evaluations of the derivatives for the typical structures
parameters. For the Al$_X$Ga$_{1-X}$As/GaAs heterostructures I
have $\beta= 0.28 \times 10^{11}$\,(meV\,cm$^2$)$^{-1}$ and $d =
100$\,nm. Hence I get $e^2\beta /C_0 \approx 46$. This means that
the LL shift becomes 47 times faster in the magnetic field at the
integer filling factor. As for the subband energy $E_0$ it also
has the large derivative that signifies a rapid increase with the
magnetic field. Thus the central energy $E_{\rm{LL}}$ demonstrates
the step-like behavior while $E_0$ has a saw-tooth oscillation on
the magnetic field (see Fig.~\ref{fig:oscil}). This is
qualitatively consistent with experimental observations and
numerical calculations.\cite{Eaves}

The strong field dependencies of the levels $E_0$ and
$E_{\rm{LL}}$ can explain an extraordinarily field sensitivity of
the SCA features.\cite{Tessmer1} Actually let me imagine an
inhomogeneous variation of the subband and the PFLL energies in
the 2DES. This means different local Fermi densities of states
$G_F$ that corresponds to different derivatives
$dE_{\rm{LL}}/d\varepsilon_c$ and $dE_0/d\varepsilon_c$ in
accordance with Eqs.~(\ref{derE0}) and (\ref{derELL}). Due to the
difference of the derivatives PFLL broadening or a dispersion of a
$E_{\rm{LL}}$ distribution changes with a magnetic field
variation. The grater are values of the derivatives
$dE_{\rm{LL}}/d\varepsilon_c$ the grater is variation of the
dispersion. Thus at low values of $G_F$, namely, near integer LL's
filling the $E_{\rm{LL}}$ dispersion should be most sensitive to
the magnetic field variation. Moreover since the derivative
$dE_{\rm{LL}}/d\varepsilon_c$ increases at the $C_0$ decrease, the
closer is the reservoir to the 2DES (the lesser is $d$) the lesser
are the derivatives $dE_{\rm{LL}}/d\varepsilon_c$ [see
Eq.~(\ref{derELL})] and the lesser is the sensitivity of the $G_F$
and the $E_{\rm{LL}}$ dispersions and the SCA image contrast. This
was observed in SCA images measured on a 2DES tunnel coupled to a
$n^+$ reservoir.\cite{Tessmer2} For instance, one can compare
Figure~3 in Ref.~\onlinecite{Tessmer2} and Figure~3 in
Ref.~\onlinecite{Tessmer1}\footnote{A direct comparison of the
figures is not eligible because different values were measured in
in-phase and out-of-phase components of SCA signal. In
particularly the data shown in Fig.~3 of
Ref.~\onlinecite{Tessmer1} correspond to the in-phase SCA signal
that is proportional to the local compressibility of the 2DES
while those shown in Ref.~\onlinecite{Tessmer2} are proportional
to a tunnel resistance of a barrier separated the 2DES and the
$n^+$ reservoir. The direct comparison is possible only after
finding of an appropriate explanation of the features observed in
the SCA images. But here I would like to pay attention only on a
sensitivity of the features to the magnetic field variation and
compare it with that of $E_0$ and $G_F$. This comparison can be
justified by that the 2DES compressibility is proportional to
$G_F$ at least in the single particle approximation and the tunnel
barrier transparency is determined by $E_0$.}. In former case the
image contrast changed markedly at the magnetic field variation of
$\delta B_1 = 1$\,kOe and in the last case it is of $\delta B_2 =
50$\,Oe. According to Eq.~(\ref{ELLjump}) at low values of $G_F$
$dE_{\rm{LL}}/d\varepsilon_c$ is inverse proportional to $C_0$
that can be estimated from parameters of investigated structures.
Thus in the case of Ref.~\onlinecite{Tessmer2} the tunnel distance
is 40\,nm that corresponds to $C_0 = \kappa/d =
2.8\times10^{-3}$\,F/m$^2$. The parameters of
Ref.~\onlinecite{Tessmer1} were thoroughly considered in
Ref.~\onlinecite{Tessmer} where a $C_{\rm{stray}}$ can be
considered as a capacity between the 2DES and the ohmic contact or
the reservoir. Authors had also evaluated the stray capacity as
1\,fF. At this case a charging area of the 2DES is a disk of $L =
90$ nm diameter under the tip. Hence the equivalent capacity
$C_{0s}$ can be estimated as $C_{0s} = C_{\rm{stray}}/L^2 =
0.12$\,F/m$^2$. By this means one can compare the relation of the
magnetic field variations $\delta B_1/\delta B_2 = 20$ and that of
the LL centers derivatives, which can be estimated as a relation
of the capacities $C_{0s}/C_0 \approx 43$ in accordance with
Eq.~(\ref{ELLjump}). One can see that the relations are of the
same order and the model proposed in this paper can be applied for
description of the local probe experiments. However to describe
the SCA features in details some improvement of the model is
necessary. First of all the in-plane charge transfer should be
considered. In this case the $C_0$ dependence upon $G_F$ should be
found. Many-body effects also can be included in the model. The
derivative $dE_F/dn$ therewith should be locally found and
substituted in the place of $G_F$.

 Analytic Eqs.~(\ref{derELL}) and (\ref{derE0}) allow
 to determine a magnetic field
value when PFLL pinning starts. This question is not simple
because according to Eq.~(\ref{derELL}) the derivative
$dE_{\rm{LL}}/d\varepsilon_c$ is always positive. Hence there is
no a sharp LL pinning transition but some criterion can be found.
To do this let me consider a discrepancy between a LL pinning
picture and a LL behavior in a closed 2DES with a constant
electron number. In the last case a subband energy $E_0$ is
constant and $dE_{\rm{LL}}/d\varepsilon_c = N + 1/2$. Under the LL
pinning condition $E_{\rm{LL}}$ is constant and Eq.~(\ref{pinE0})
is valid. Thus it seems to be reasonable to choose
$dE_{\rm{LL}}/d\varepsilon_c = - dE_0/d\varepsilon_c = (N +
1/2)/2$ as a starting point for LL pinning. Hence according to
Eq.~(\ref{derE0}) and one can get the following:
\begin{equation}\label{GFstart}
    \frac{e^2n+\varepsilon_{c0}C_0(N+1/2)}{\varepsilon_{c0}e^2G_F+\varepsilon_{c0}C_0}=\frac{N+1/2}{2},
\end{equation}
where $\varepsilon_{c0}$ is the cyclotron energy at the LL pinning
start. Since I'm interesting in the starting point I should
consider $G_F$ value in the LL center, i.e., $G_F =
2\beta\varepsilon_{c0}^{1/2}/\pi\gamma$ in accordance with
Eq.(\ref{spectral}). Also taking in to account that $N + 1/2 =
\nu/2 - 1 \approx n/\beta\varepsilon_{c0}$ for the half populated
PFLL the start cyclotron energy $\varepsilon_{c0}$ can be found as
\begin{equation}\label{ec0}
    \varepsilon_{c0}=\pi^2\gamma^2\left(1+\frac{C_0}{2\beta
    e^2}\right)^2.
\end{equation}
It worth to note that this value is one order larger than
cyclotron energy of the LLs resolution, which is determined from
$\varepsilon_{c1}\approx\Gamma$ as $\varepsilon_{c1} \approx
\gamma^2$.

To finish the section it will be pertinent to discuss real
structure parameters and magnetic fields to those the developed
model is applicable. First of all one should satisfy $N \gg 1$ and
$\varepsilon_c \gg \Gamma$ conditions. This definitely requires
quite high mobility and concentration of electrons in a 2DES. Let
me suppose a ten times larger than one as a much more than one. So
for a quite usual mobility of the 2DES as $\eta =
100$\,m$^2$/(V$\times$s) one can resolve LLs at $B = 1/\eta =
0.01$\,T. Since according to Eq.~(\ref{ec0}) PFLL pinning will
start at $B = B_0 \approx 0.1$\,T. In this case one should provide
at least ten LLs populated that corresponds to $E_F \approx
2$\,meV or $n \approx 6 \times 10^{10}$\,cm$^{-2}$. It is worth
noting the larger is the electron concentration the better is the
model applicability. Now one should provide homogeneous LL
broadening that is to locate a metal at the distance $d \ll \ell$
from the 2DES plane. At $B = B_0$ the magnetic length $\ell$ is
about 250\,nm. Thus $d = 25$\,nm will be sufficient to assure
homogeneous broadening of the LLs up to $B = 10$\,T. At the last
step one need to justify inequality~(\ref{width}). This means that
QW width should be about 3\,nm.

\section{\label{sec:concl}CONCLUSIONS}
In conclusion it is noteworthy that some results can be expanded
on a closed 2DES with a constant electron concentration $n$. In
this case it would be more relevant to say about a Fermi level
capture by the PFLL. Condition of the capture can be derived in
similar way like that of LL pinning. It can be obtained from
Eq.~(\ref{pincon}) setting $C_0=0$ that is equivalent to an
infinitely far reservoir. The same setting should be apply to get
a start of the Fermi level capture from Eq.~(\ref{ec0}).

Summarizing thermodynamic equilibrium between a 2DES and a 3D
electron system in a quantizing magnetic field has been considered
in the case of the homogeneous broadened Landau levels. Parameters
of the proposed model have allowed to consider analytically such
effects as PFLL pinning and the subband energy oscillations in the
2DES. A condition and a starting point of PFLL pinning have been
defined and a starting value of a cyclotron energy has been
determined. Moreover found dependencies of $E_{\rm{LL}}$ and $E_0$
up on $C_0$ allow to describe qualitatively the extraordinary
sensitivity of the SCA features observed recently in local probe
experiments. All results are in a good qualitative agreement with
experimental data.

\begin{acknowledgments}

I would like to thank V. A. Volkov and E. E.
Takhtamirov for their sincere interest and valuable discussions.
This work was supported by RFBR (grant  04-02-16870).

\end{acknowledgments}

\bibliography{pinning1}

\end{document}